\documentclass[pra,aps,showpacs,nofootinbib,preprint]{revtex4}
\usepackage[dvips]{graphicx} 
\newcommand{\fix}{\Phi(\mathbf{x})}
\newcommand{\fiLx}{\Phi_{\rm L}(\mathbf{x})}

\newcommand{\fik}{\Phi(\mathbf{k})}
\newcommand{\fiLk}{\Phi_{\rm L}(\mathbf{k})}
\newcommand{\fiLkone}{\Phi_{\rm L}(\mathbf{k_1})}
\newcommand{\fiLktwo}{\Phi_{\rm L}(\mathbf{k_2})}

\newcommand{\fiNLk}{\Phi_{\rm NL}(\mathbf{k})}

\newcommand{\fiNLkthree}{\Phi_{\rm NL}(\mathbf{k_3})}

\newcommand{\kernel}{f_{\rm NL} (\mathbf{k_1},\mathbf{k_2},\mathbf{k_3})}

\newcommand{\dirackonektwokthree}{\delta^{(3)}\,(\mathbf{k_1+k_2+k_3})}

\newcommand{\beq}{\begin{equation}}
\newcommand{\eeq}{\end{equation}}
\newcommand{\beqarr}{\begin{eqnarray}}
\newcommand{\eeqarr}{\end{eqnarray}}

\newcommand{\angk}{\hat{k}}
\newcommand{\angn}{\hat{n}}

\newcommand{\tfnow}{\Delta_\ell(k,\tau_0)}
\newcommand{\tf}{\Delta_\ell(k)}
\newcommand{\tfone}{\Delta_{\ell_1}(k_1)}
\newcommand{\tftwo}{\Delta_{\ell_2}(k_2)}
\newcommand{\tfthree}{\Delta_{\ell_3}(k_3)}

\newcommand{\alm}{a_{\ell m}}
\newcommand{\almL}{a_{\ell m}^{\rm L}}
\newcommand{\almNL}{a_{\ell m}^{\rm NL}}
\newcommand{\almone}{a_{\ell_1 m_1}}
\newcommand{\almLone}{a_{\ell_1 m_1}^{\rm L}}

\newcommand{\almtwo}{a_{\ell_2 m_2}}
\newcommand{\almLtwo}{a_{\ell_2 m_2}^{\rm L}}

\newcommand{\almthree}{a_{\ell_3 m_3}}

\newcommand{\almNLthree}{a_{\ell_3 m_3}^{\rm NL}}

\newcommand{\YLMstar}{Y_{L M}^*}
\newcommand{\Ylmstar}{Y_{\ell m}^*}
\newcommand{\Ylmstarone}{Y_{\ell_1 m_1}^*}
\newcommand{\Ylmstartwo}{Y_{\ell_2 m_2}^*}
\newcommand{\Ylmstarthree}{Y_{\ell_3 m_3}^*}

\newcommand{\YLM}{Y_{L M}}
\newcommand{\Ylm}{Y_{\ell m}}

\newcommand{\Ylmfour}{Y_{\ell_1^\prime m_1^\prime}}
\newcommand{\Ylmfive}{Y_{\ell_2^\prime m_2^\prime}}
\newcommand{\Ylmsix}{Y_{\ell_3^\prime m_3^\prime}}

\newcommand{\jlfourone}{j_{\ell_1^\prime}(k_1 r)}
\newcommand{\jlfivetwo}{j_{\ell_2^\prime}(k_2 r)}
\newcommand{\jlsixthree}{j_{\ell_3^\prime}(k_3 r)}

\newcommand{\Gaunt}{\mathcal{G}_{\ell_1^\prime \, \ell_2^\prime \, \ell_3^\prime}^{m_1^\prime m_2^\prime m_3^\prime}}
\newcommand{\Gaunttwo}{\mathcal{G}_{\ell_1^\prime \, \ell_2^\prime \, \ell_3}^{m_1^\prime m_2^\prime m_3}}

\newcommand{\Gauntstarone}{\mathcal{G}_{\ell_1 \, L \,\, \ell_1^\prime}^{-m_1 M m_1^\prime}}
\newcommand{\Gauntstartwo}{\mathcal{G}_{\ell_2^\prime \, \ell_2 \, L}^{-m_2^\prime m_2 M}}

\newcommand{\dangn}{d \angn}

\newcommand{\dkone}{d^3 k_1}
\newcommand{\dktwo}{d^3 k_2}
\newcommand{\dkthree}{d^3 k_3}

\newcommand{\planewave}{e^{i\mathbf{k \cdot x}}}
\newcommand{\dallkfourier}{\frac{\dkone}{(2\pi)^3}\frac{\dktwo}{(2\pi)^3}\frac{\dkthree}{(2\pi)^3}}

\newcommand{\Bis}{B_{\ell_1 \ell_2 \ell_3}^{m_1 m_2 m_3}}
\newcommand{\Avbis}{B_{\ell_1 \ell_2 \ell_3}}

\newcommand{\los}{\mathcal{L}_{\ell_3 \ell_1 \ell_2}^{L \, 
\ell_1^\prime \ell_2^\prime}(r)}
\newcommand{\loszero}{\mathcal{L}_{\ell_3 \ell_1 \ell_2}^{0 \, 
\ell_1^\prime \ell_2^\prime}(r)}
\newcommand{\losone}{\mathcal{L}_{\ell_3 \ell_1 \ell_2}^{1 \, 
\ell_1^\prime \ell_2^\prime}(r)}
\newcommand{\lostwo}{\mathcal{L}_{\ell_3 \ell_1 \ell_2}^{2 \, 
\ell_1^\prime \ell_2^\prime}(r)}
\newcommand{\losfNL}{\mathcal{L}_{\ell_3 \ell_1 \ell_2}^{0 \, 
\ell_1 \ell_2}(r)}

\begin{document}

\preprint{FERMILAB-PUB-05-386-T}

\title{Testing Primordial Non-Gaussianity in CMB Anisotropies}  

\author{Michele Liguori}
\affiliation{Particle Astrophysics Center, Fermi
        National Accelerator Laboratory, Batavia, Illinois \ 60510-0500, 
USA \\
Dipartimento di Fisica ``G.\ Galilei'' Universit\`{a} di 
Padova, INFN Sezione di Padova, via Marzolo 8, I-35131 Padova, Italy}

\author{Frode K. Hansen}
\affiliation{Institute of Theoretical Astrophysics, University of Oslo, 
P.O. Box 1029 Blindern, 0315 Oslo, Norway}

\author{Eiichiro Komatsu}
\affiliation{Department of Astronomy, University of Texas at Austin,
1 University Station, C1400, Austin, TX 78712, USA}

\author{Sabino Matarrese}
\affiliation{Dipartimento di Fisica ``G.\ Galilei'' Universit\`{a} di 
Padova, INFN Sezione di Padova, via Marzolo 8, I-35131 Padova, Italy}

\author{Antonio Riotto}
\affiliation{INFN Sezione di Padova, via Marzolo 8, I-35131 Padova, Italy}

%\date{\today}

\begin{abstract}
\noindent

Recent second-order perturbation computations have provided an accurate
prediction for the primordial gravitational potential, $\fix$,
in scenarios in which cosmological perturbations are generated either
during or after inflation. This enables us to make realistic
predictions for a non-Gaussian part of $\fix$, which is generically
written in momentum space as a double convolution of its Gaussian part
with a
suitable kernel, $f_{\rm NL}(\mathbf{k}_1,\mathbf{k}_2)$.
This kernel defines the amplitude and
angular structure
of the non-Gaussian signals and originates from the
evolution of
second-order
perturbations after the generation of the curvature
perturbation. We derive a generic
formula for the CMB
angular bispectrum with arbitrary $f_{\rm NL}(\mathbf{k}_1,\mathbf{k}_2)$,
 and examine the
detectability of the
primordial non-Gaussian signals from various scenarios such as
single-field inflation,
inhomogeneous reheating, and curvaton scenarios. Our results show that
in the standard slow-roll inflation scenario
the signal actually comes from the momentum-dependent part of
$f_{\rm NL}(\mathbf{k}_1,\mathbf{k}_2)$, and thus
it is important to include the momentum dependence in the data analysis.
In the other scenarios the primordial non-Gaussianity is
comparable to or larger
than these post-inflationary effects.
We find that {\sl WMAP} cannot detect non-Gaussian signals generated by
these models.
Numerical calculations for $l>500$ are still computationally expensive,
and we are not yet able to extend our calculations to {\sl Planck}'s
angular
resolution; however,
there is an encouraging trend which shows that {\sl Planck} may be able to
detect these non-Gaussian signals.
\end{abstract} 

\pacs{98.80.Cq, 95.35.+d, 4.62.+v}

\maketitle
%%%%%%%%%%%%%%%%%%%%%%%%%%%%%%%%%%%%%%
%%%%%%%%%%%%%%%%%%%%%%%%%%%%%%%%%%%%%%
\section{Introduction}
%%%%%%%%%%%%%%%%%%%%%%%%%%%%%%%%%%%%%%
%%%%%%%%%%%%%%%%%%%%%%%%%%%%%%%%%%%%%%
\noindent
Inflation is a building-block of the standard model of modern cosmology. 
It is widely believed that there was an early stage
in the history of the Universe -- 
before the epoch of primordial nucleosynthesis -- when the 
 expansion rate of the Universe was accelerated. Such a period of 
cosmological inflation can be attained if the energy density of 
the Universe is dominated by the vacuum energy density associated with the 
potential of a scalar field, called the inflaton \cite{lrreview}. 
Inflation has become so popular also because 
of another compelling feature. It provides a causal mechanism 
for the production of the first density perturbations in the early 
Universe which are the seeds for the Large-Scale Structure (LSS) 
of the Universe and for the Cosmic Microwave Background (CMB) 
temperature and polarization anisotropies that we observe today. 
% In fact inflation has become the dominant paradigm to understand the 
% initial conditions for structure formation and CMB anisotropies. 
In the inflationary picture, 
primordial density and gravity-wave fluctuations were
created from quantum fluctuations and then left 
%``redshifted'' out of 
the horizon during an early period of superluminal expansion of the Universe, 
%where they are ``frozen''. 
with the amplitude ``frozen-in''.
Perturbations at the surface of last scattering are observable as temperature 
and polarization anisotropies in the CMB.
%The last and most impressive confirmation of 
The inflationary paradigm has been 
%recently provided 
tested carefully by the data of the Wilkinson Microwave Anisotropy Probe 
({\sl WMAP}) mission~\cite{bennett/etal:2003b}.
The {\sl WMAP} collaboration has produced a full-sky map of the angular 
variations of the CMB with unprecedented accuracy.
The {\sl WMAP} data confirm the inflationary mechanism as responsible for the
generation of curvature (adiabatic) superhorizon fluctuations \cite{ex}. 
Since the primordial cosmological perturbations are tiny, the generation
and evolution of fluctuations during inflation has been studied 
within linear perturbation theory. Within this approach, 
the primordial density perturbation field is a
Gaussian random field; in other words, its Fourier components
are uncorrelated and have random phases. 
Despite the simplicity of the inflationary paradigm, the mechanism
by which  cosmological adiabatic perturbations are generated is not
yet fully established. In the standard slow-roll scenario associated
with one-single field models of inflation, the observed density 
perturbations are due to fluctuations of the inflaton field itself when it
slowly rolls down along its potential. 
When inflation ends, the inflaton oscillates about the minimum of its
potential and decays, thereby reheating the Universe. 
As a result of the fluctuations
each region of the Universe goes through the same history but at slightly
different times. The final temperature anisotropies are caused by 
inflation lasting for different amounts of time in 
different regions of the Universe
leading to adiabatic perturbations \cite{lrreview}.

An alternative to the standard scenario is represented by the curvaton 
mechanism~\cite{Mollerach,curvaton1,LW2,curvaton3,LUW}
where the final curvature perturbations are produced from an initial 
isocurvature perturbation associated with the quantum fluctuations of a 
light scalar field (other than the inflaton), the curvaton, whose energy 
density is negligible during inflation. The 
curvaton isocurvature perturbations are transformed into adiabatic
ones when the curvaton decays into radiation, much after the end 
of inflation. 

Recently, other mechanisms for the generation of cosmological
perturbations have been proposed, see \cite{komreview} for a 
comprehensive review. 
For instance, the inhomogeneous reheating scenario \cite{gamma} 
acts during the reheating stage after inflation if superhorizon 
spatial fluctuations in the 
decay rate of the inflaton field are induced during inflation, causing 
adiabatic perturbations in the final reheating temperature
in different regions of the Universe. Alternatively, curvature perturbations
may be created because of the presence of broken symmetries during inflation
\cite{val}.

Testing the Gaussianity of the primordial fluctuations provides
a powerful tool to discriminate between different scenarios for the
generation of the cosmological perturbations which would be 
indistinguishable otherwise \cite{komreview}.
Non-Gaussianity 
is a deviation from a pure Gaussian statistics, {\it i.e.}, the
presence of higher-order connected correlation functions of CMB anisotropies. 
The angular $n$-point correlation function
is a simple statistic characterizing a clustering pattern of
fluctuations on the CMB. 
If the fluctuations are Gaussian, then the two-point correlation function
specifies all the statistical properties of higher-order correlation
functions, for
the two-point correlation function is the only parameter in a Gaussian 
distribution. 
If it is not Gaussian, then we need higher-order correlation 
functions to determine the statistical properties.
For instance, a non-vanishing
three-point function of scalar perturbations, or its Fourier transform,
the bispectrum, is an indicator of non-Gaussian features in the
cosmological perturbations. 
The importance of the bispectrum comes from the fact that it represents
the lowest order statistics able to distinguish non-Gaussian from Gaussian
perturbations.
An accurate calculation of the primordial bispectrum of cosmological
perturbations has become an extremely important issue, as a number of
present and future experiments, such as {\sl WMAP} and {\sl Planck}, 
will allow us to constrain or detect non-Gaussianity of CMB anisotropy 
with high precision.

In order to compute and keep track of the 
non-Gaussianity of the cosmological perturbations throughout the different 
stages of the evolution of the Universe, one has to perform a perturbation 
around the homogeneous background up to second order. 
Recent studies have been able to characterize the level of non-Gaussianity
predicted in the various scenarios for the generation of the
cosmological perturbations 
\cite{komreview,acqua,maldacena,BMR2,BMR3,BMR4,BMR5,BMR6,lythetal}. 

On large scales the second-order, gauge-invariant expression for 
the temperature 
anisotropies reads \cite{BMR5,komreview,BMR6}
\begin{equation}
\label{main}
\frac{\Delta T_2}{T}=
\frac{1}{18} \Phi_{\rm L}^2 
-\frac{{\mathcal K}}{10}-\frac{1}{10} \left( \zeta_2-
2 \zeta_{\rm L}^2 \right)\, , 
\end{equation} 
where  $\Phi_{\rm L}$ represents the gauge-invariant 
gravitational potential
at linear order, $\zeta_{\rm L}$ is the linear gauge-invariant
comoving curvature perturbation, $\zeta_2$ is the second-order
gauge-invariant comoving curvature perturbation, and 
%we have defined 
\begin{equation}
{\mathcal K}\equiv 10 \nabla^{-4} \partial_i \partial^j 
 (\partial^i \Phi_{\rm L} \partial_j
\Phi_{\rm L}) -\nabla^{-2} 
\left( \frac{10}{3} \partial^i  \Phi_{\rm L} \partial_i   \Phi_{\rm L}\right).
\end{equation} 
It shows that there are two contributions to the final 
nonlinearity in the 
large-scale temperature anisotropies. 
The third term, 
%contribution 
$\left(\zeta_{2}-
2 \zeta_{\rm L}^2\right)$, 
comes from the ``primordial'' conditions set during or after inflation. 
They are encoded in 
the curvature perturbation $\zeta$ which remains constant 
once it has been generated. 
The remaining part of Eq.~(\ref{main}) describes the post-inflation
processing of the primordial non-Gaussian signal due to the nonlinear 
gravitational dynamics, including also second-order corrections at last 
scattering to the Sachs-Wolfe effect. 
Thus, the expression in Eq.~(\ref{main}) allows us to separate the 
primordial contribution to non-Gaussianity from that arising 
after inflation. 

While the nonlinear evolution after inflation is the same in 
each scenario, the primordial term will depend on
the particular mechanism generating the perturbations.
We may parametrize the primordial 
non-Gaussianity in the terms of the conserved curvature 
perturbation (in the radiation or matter dominated epochs)
$\zeta_2=2a_{\rm NL}\left(\zeta_{\rm L}\right)^2$, 
where $a_{\rm NL}$ depends on the physics of a given scenario. 
Within the standard scenario 
where cosmological perturbations are due to the inflaton the 
primordial contribution to the non-Gaussianity is given by 
$a_{\rm NL}=1-\frac{1}{4} (n_{\zeta}-1)$~\cite{acqua,maldacena,BMR2}, 
where the spectral index is expressed in terms of the usual 
slow-roll parameters as $n_{\zeta}-1=-6 \epsilon + 
2 \eta$~\cite{lrreview}. 
In the curvaton case $a_{\rm NL}=(3/4r)-r/2$, where 
$r \approx (\rho_\sigma/\rho)_{\rm D}$ is the relative   
curvaton contribution to the total energy density at curvaton 
decay~\cite{komreview}. In the minimal picture for the inhomogeneous 
reheating scenario, $a_{\rm NL}=1/4$.
 
>From Eq.~(\ref{main}) one can extract the nonlinearity parameter 
$f_{\rm NL}$ which is usually adopted to phenomenologically 
parametrize the non-Gaussianity level of cosmological perturbations and 
has become the standard quantity to be observationally
constrained by CMB experiments~\cite{ks,k}.
The comparison between our expression [Eq.~(\ref{main})] and 
that in the previous work \cite{ks,k} can be made through the 
Sachs-Wolfe formula, $\Delta T/T=-(1/3)\Phi$,  
where $\Phi$ is Bardeen's gauge-invariant potential, which is 
conventionally expanded as (up to a constant offset, which only 
affects the temperature monopole)

\begin{equation}
\Phi = \Phi_{\rm L} + f_{\rm NL} \star \Phi_{\rm L}^2\, .\end{equation} 
Here the $\star$-product (convolution) makes explicit the fact that the nonlinearity 
parameter has a non-trivial scale dependence~\cite{komreview}. 
Therefore, using  $\zeta_{\rm L}=-(5/3) 
\Phi_{\rm L}$ during matter domination,
from Eq.~(\ref{main}) we may define the nonlinearity parameter in 
momentum space  
\begin{equation}
\label{eqn:f_NL}
f_{\rm NL}({\bf k}_1,{\bf k}_2)
=-\left[ \frac{5}{3} \left(1-a_{\rm NL} \right) 
+\frac{1}{6}-\frac{3}{10} {\mathcal K} 
\right]\, ,
\end{equation}
where 
\begin{equation}
{\mathcal K}=
\frac{10\, ({\bf k}_1 \cdot {\bf k}_3)({\bf k}_2 \cdot {\bf k}_3)}{k^4_3} 
-\frac{10({\bf k}_1 \cdot {\bf k}_2)}{3k^2_3},
\end{equation}
and ${\bf k}_3+{\bf k}_1+{\bf k_2}=0$ and $k_3=\left|{\bf k}_3\right|$\footnote{The formula
(\ref{eqn:f_NL}) already accounts for an additional nonlinear effect 
entering in the CMB angular $3$-point function from the angular 
averaging performed with a perturbed line-element 
implying a $+1$ shift 
in $f_{\rm NL}$.}. Notice that in 
the ``squeezed'' limit first discussed by Maldacena~\cite{maldacena}, 
where one of the wavenumbers is much smaller than the other two, 
\emph{e.g.} $k_1 \ll k_{2,3}$, the momentum dependence
of the kernel disappears. 

The fact that the nonlinearity parameter has a scale (momentum) 
dependence, that is that $f_{\rm NL}$ is not simply
a number, may call for a reanalysis of the  
tests performed so far of the non-Gaussianity 
in the primordial cosmological perturbations \cite{ks,k}. This is because
previous studies have been done when theoretical predictions for the
nonlinearity parameters in the various scenarios (including the standard
case in which perturbations are generated by the inflaton field)
were not available and therefore $f_{\rm NL}$ 
was assumed phenomenologically to be a constant. 

The observational capability of determining the nonlinearity
parameter $f_{\rm NL}$ is the subject of a long project of which
this paper represents the first step.  
Starting from a generic expression for the gravitational 
potential, we first derive the generic expression for the 
primary CMB angular bispectrum. This formula generalizes the one 
provided by Komatsu and Spergel \cite{ks} who worked with a 
constant $f_{\rm NL}$ in momentum space.
We then estimate the expected signal-to-noise ratio
for detecting primary non-Gaussianity at {\sl WMAP} angular resolution. 
While we show that the primary non-Gaussian 
signal generated in standard scenarios of inflation cannot be detected by 
{\sl WMAP}, our predicted signal-to-noise ratio shows a trend which, if maintained at 
higher angular resolution, should allow detection of the non-Gaussian signals  
 by the future {\sl Planck} 
mission even in the standard single-field scenario of inflation -- in this case,
$f_{\rm NL}$ is dominated by the post-inflationary evolution, rather than the 
primordial contribution from inflation.

The paper is organized as follows. In section \ref{sec:def} we give some basic 
definitions and we compute analytically the 
CMB angular bispectrum arising from a primordial potential of the kind
described by equation (\ref{eqn:f_NL}); in section \ref{sec:results} we 
present our numerical predictions for the primary angular 
bispectrum, and discuss detectability with the current and future experiments; 
section \ref{sec:conclusions} contains our concluding remarks.

\section{The CMB angular bispectrum}\label{sec:def}

\subsection{Basics}

%Let us start with some definitions. 
The CMB angular bispectrum is defined by

\beq\label{eqn:bispectrumdefinition}
\Bis = \left\langle \almone \almtwo \almthree \right\rangle \; \; ,
\eeq
where we have expanded the observed CMB temperature 
fluctuations into spherical harmonics, and 
we have defined the multipoles $a_{\ell m}$

\beq
\alm = \int d^2 \angn \frac{\Delta T \left(\angn\right)}{T} \Ylmstar \; .
\eeq
We find it convenient to split 
the multipoles $\alm$ into a Gaussian part $\almL$ and a non-Gaussian part
$\almNL$:

\beq
\alm = \almL + \almNL \; .
\eeq
By ignoring second-order terms in $\almNL$, we obtain

\beq\label{eqn:ourbisdef}
\Bis = \left\langle \almLone \almLtwo \almNLthree \right\rangle
+ 
\left(\begin{array}{c} 
\ell_3 \leftrightarrow \ell_1 \\
\ell_2 \leftrightarrow \ell_3 
\end{array}\right)
+
\left(\begin{array}{c} 
\ell_3 \leftrightarrow \ell_2 \\
\ell_1 \leftrightarrow \ell_3 
\end{array}\right) \; .
\eeq
The rotational invariance of the CMB sky implies that $\Bis$ 
can always be decomposed as

\beq\label{eqn:angavbis}
\Bis =  \left( \begin{array}{ccc} 
               \ell_1 & \ell_2 & \ell_3 \\
                 m_1      &  m_2     &  m_3    
               \end{array} \right)
        \Avbis \; . 
\eeq
Where $\Avbis$ is the angle-averaged bispectrum and the matrix 
is the Wigner $3j$ symbol. The presence of the Wigner $3j$ symbol
ensures that the bispectrum satisfies the selection rules, 
$m_1 + m_2 + m_3 = 0$, $\ell_1 + \ell_2 + \ell_3 =$ even, and the 
triangle conditions, $|\ell_i - \ell_j| \leq \ell_k \leq \ell_i 
+ \ell_j$ for all permutation of indices $i,j,k$.

As we have mentioned in the Introduction, in the various scenarios for the
generation of the cosmological perturbations, 
the non-Gaussian part of the primordial gravitational potential 
can be expressed in Fourier space as a double convolution, 

\beq\label{eqn:NGpotential}
\Phi_{\rm NL}(\mathbf{k_3}) = \frac{1}{(2 \pi)^3} \int \! d^3k_1\,d^3k_2 
\,\delta^{(3)}\,(\mathbf{k_1+k_2-k_3})
\,\Phi_{\rm L}(\mathbf{k_1})\,\Phi_{\rm L}(\mathbf{k_2}) 
f_{\rm NL} (\mathbf{k_1},\mathbf{k_2},\mathbf{k_3})
\; ,
\eeq
where $\Phi_{\rm L}({\bf k})$ 
is a Gaussian random field representing the Gaussian
part of the primordial potential; the kernel, $\kernel$, 
in equation (\ref{eqn:NGpotential}) can be written, without loss
of generality, as:

\beq\label{eqn:kernelexpression}
\kernel = \sum_{n=0}^N \frac{c_n(k_1,k_2) 
(\angk_1 \cdot \angk_2)^n}{k_3^{2n}} \; ;
\eeq
in the following we are going to expand 
$f_{\rm NL}(\mathbf{k_1},\mathbf{k_2},\mathbf{k_3})$ in Legendre polynomials
in terms of the angle between $\mathbf{k_1}$ and $\mathbf{k_2}$\footnote{
In equation 
(\ref{eqn:f_NL}) we defined the kernel as a function of $\mathbf{k_1}$ and 
$\mathbf{k_2}$ only,
 as $\mathbf{k_3}$ was given by $\mathbf{k_3} = \mathbf{k_1} + \mathbf{k_2}$; 
nevertheless 
for our following derivation of the bispectrum we find it convenient to 
introduce the Dirac delta function $\dirackonektwokthree$ in eqn. 
(\ref{eqn:NGpotential}) and to write the convolution kernel 
 as a function of  $\mathbf{k_1}$, 
$\mathbf{k_2}$, $\mathbf{k_3}$ separately. In this way 
we can avoid 
factors of $(\angk_1 \cdot \angk_2)^n$ in the denominator of 
(\ref{eqn:kernelexpression}). Those factors would make the decomposition
of the kernel in Legendre polynomials difficult.}:

\beq\label{eqn:kernelexpansion} 
f_{\rm NL}(\mathbf{k_1},\mathbf{k_2},\mathbf{k_3})
 = \sum_{\ell=0}^{N} f_\ell(k_1,k_2,k_3) 
P_\ell(\angk_1 \cdot \angk_2) \; .
\eeq

The multipoles of the harmonic expansion of the (today observed) CMB 
temperature anisotropies are related to
the primordial potential $\fik$, the relation between the two quantities
being described by the linear radiation transfer functions, $\tfnow$:

\beq\label{eqn:phi2alm}
\alm = (-i)^\ell \int \! \frac{d^3k}{(2\pi)^3} \, \fik \, \tfnow \, 
\Ylmstar(\angk) \; ,
\eeq 
where we are evolving the primordial perturbations up to the 
present time $\tau_0$. 
In the following we 
%will omit to specify this parameter in the 
%transfer functions, by 
write simply $\tf$ instead of $\tfnow$. 

The primordial potential is the sum of a linear and a 
nonlinear part:
$\fik = \fiLk + \fiNLk$, where the non-Gaussian part is given 
by formula (\ref{eqn:NGpotential}); accordingly, we can split also 
the temperature 
fluctuation and the multipoles $\alm$ into Gaussian and non-Gaussian 
components. Our aim in the next section will be to calculate 
the CMB angular bispectrum, 
starting from the bispectrum of the primordial gravitational 
potential which is, by definition
$\left\langle \fiLkone \fiLktwo \fiNLkthree \right\rangle +$ 
cyclic permutations. 

\subsection{Analytic formula of the primary bispectrum with arbitrary kernel}
\noindent
%To begin, 
Let us first fix the notation by explicitly writing 
Eq.~(\ref{eqn:phi2alm}) 
for $\almLone$, $\almLtwo$ and $\almNLthree$:
\beqarr
\almLone & = & (4 \pi) (-i)^{\ell_1} \! \int \! \frac{\dkone}{(2 \pi)^3} \, 
\fiLkone 
	       \, \Ylmstarone(\angk_1) \, \Delta_{\ell_1}(k_1) 
\label{eqn:almonedef} \; ,\\
\almLtwo & = & (4 \pi) (-i)^{\ell_2} \! \int \! \frac{\dktwo}{(2 \pi)^3} 
\, \fiLktwo 
	       \, \Ylmstartwo(\angk_2) \, \Delta_{\ell_2}(k_2) \; . 
\label{eqn:almtwodef} \\
\almNLthree & = & (4 \pi) (-i)^{\ell_3}  \! \int \! 
\frac{\dkthree}{(2 \pi)^3} \, \fiNLkthree 
	       \, \Ylmstarthree(\angk_3) \, \Delta_{\ell_3}(k_3) \; . 
\label{eqn:almthreedef}
\eeqarr
Now, putting together Eqs. (\ref{eqn:almonedef}), (\ref{eqn:almtwodef}) and 
(\ref{eqn:almthreedef}), and using (\ref{eqn:ourbisdef}), we find
\beqarr
\Bis & = &       (4 \pi)^3 (-i)^{\ell_1 + \ell_2 + \ell_3} \int \! 
\dallkfourier   
                 \left\langle \fiLkone \fiLktwo \fiNLkthree 
\right\rangle          \nonumber \\
     & \times &  \tfone \, \tftwo \, \tfthree \, 
                 \Ylmstarone(\angk_1) \, \Ylmstartwo(\angk_2) \, 
\Ylmstarthree(\angk_3) \nonumber \\
     & + & 
       \left(\begin{array}{c} 
         \ell_3 \leftrightarrow \ell_1 \\
         \ell_2 \leftrightarrow \ell_3 
       \end{array}\right)
       +
       \left(\begin{array}{c} 
         \ell_3 \leftrightarrow \ell_2 \\
         \ell_1 \leftrightarrow \ell_3 
       \end{array}\right) \; .
\eeqarr
The component $\left\langle \fiLkone \fiLktwo \fiNLkthree 
\right\rangle$ of the $\fik$-field 
bispectrum can be easily calculated:
%. It reads
\beq
\left\langle \fiLkone \fiLktwo \fiNLkthree \right\rangle 
= 2(2\pi)^3\dirackonektwokthree
\kernel P(k_1)P(k_2),
\eeq
%If we define
%\beq
%\deltatilde = \frac{\tfthree}{k_3^{2N}}\, ,
%\eeq
and we obtain
\beqarr\label{eqn:first}
\Bis & = &       (4 \pi)^3 (-i)^{\ell_1 + \ell_2 + \ell_3} \int 
\! \dallkfourier   
                 2(2\pi)^3\dirackonektwokthree 
%k_3^{2N}
\kernel                               \nonumber \\
     & \times &  P(k_1)P(k_2) \tfone \, \tftwo \, \tfthree \, %\deltatilde \, 
                 \Ylmstarone(\angk_1) \, \Ylmstartwo(\angk_2) \, 
\Ylmstarthree(\angk_3)  \nonumber \\
     & + & 
       \left(\begin{array}{c} 
         \ell_3 \leftrightarrow \ell_1 \\
         \ell_2 \leftrightarrow \ell_3 
       \end{array}\right)
       +
       \left(\begin{array}{c} 
         \ell_3 \leftrightarrow \ell_2 \\
         \ell_1 \leftrightarrow \ell_3 
       \end{array}\right) \; .
\eeqarr
The Dirac delta function $\dirackonektwokthree$ can now be
written as
\beq
\dirackonektwokthree = \int \frac{d^3r}{(2 \pi)^3} 
e^{i \mathbf{k_1 \cdot r}} e^{i \mathbf{k_2 \cdot r}}
 e^{i \mathbf{k_3 \cdot r}} \; , 
\eeq
and the plane waves can be expanded according to the Rayleigh formula
\beq\label{eqn:raileygh}
\planewave = (4 \pi) \sum_\ell \sum_m (i)^\ell j_\ell(kx) \Ylm(\angk) 
\Ylmstar(\hat{x}) \; .
\eeq 
In this way we can make the substitution
\beqarr\label{eqn:diracexp}
\lefteqn
{\!\!\!\!\!\!\!\!\!\!\!\!\!\!\!\!
\!\!\!\!\!\!\!\!\!\!\!\!\!\!\!\!\!\!\!\!\!\!\!\!\!\!\!\!\!\!\!\!\!\!\!
\dirackonektwokthree = 8 \sum_{\ell_1^\prime \ell_2^\prime \ell_3^\prime} \, \sum_{m_1^\prime m_2^\prime m_3^\prime} 
(i)^{\ell_1^\prime+\ell_2^\prime+\ell_3^\prime} 
\, \mathcal{G}_{\ell_1^\prime \ell_2^\prime \ell_3^\prime}^{m_1^\prime m_2^\prime m_3^\prime} \, \Ylmfour(\angk_1) 
\Ylmfive(\angk_2)
\Ylmsix(\angk_3) 
} \nonumber \\
&\times& \int dr r^2 \jlfourone \jlfivetwo \jlsixthree \; ,
\eeqarr
where we have introduced the Gaunt integral $\mathcal{G}_{\ell_1^\prime 
\ell_2^\prime \ell_3^\prime}^{m_1^\prime m_2^\prime m_3^\prime}$, 
defined by
\beqarr\label{eqn:gauntintegral}
\Gaunt & = & \int \dangn \Ylmfour(\angn) \Ylmfive(\angn) 
\Ylmsix(\angn) \; \\  
       & = & \sqrt{\frac{(2 \ell_1^\prime +1)(2 \ell_2^\prime + 1)(2 \ell_3^\prime + 1)}{4 \pi}}
             \left( \begin{array}{ccc} 
	     \ell_1^\prime & \ell_2^\prime & \ell_3^\prime \\
	       0      &  0     &  0    
	           \end{array} \right)
             \left( \begin{array}{ccc} 
             \ell_1^\prime & \ell_2^\prime & \ell_3^\prime \\
               m_1^\prime      &  m_2^\prime     &  m_3^\prime    
             \end{array} \right) \nonumber \, .
\eeqarr
The kernel $\kernel$ can be expanded  
in spherical harmonics as well: Eq.~(\ref{eqn:kernelexpansion}),
together with the addition theorem of spherical harmonics,
\beq\label{eqn:additiontheorem}
P_L (\angk_1 \cdot \angk_2) =  
\frac{4 \pi}{2\ell+1} \sum_{M = -L}^{L} \YLM(\angk_1) 
\YLMstar(\angk_2) \; ,
\eeq
finally yields
\beqarr\label{eqn:kernelexp}
%k_3^{2N} 
\kernel & = & \sum_{L=0}^{N} f_L(k_1,k_2,k_3) 
                           P_L(\angk_1\cdot \angk_2) \nonumber \\ 
                     & = &  \sum_{L=0}^{N} \frac{4 \pi}{2L+1}
                            f_L(k_1,k_2,k_3) \sum_{M = 
-L}^{L} \YLM(\angk_1) 
                            \YLMstar(\angk_2) \; .
\eeqarr
Now, splitting the integral on the right-hand side of 
Eq.~(\ref{eqn:first}) into a radial
and an angular part, and considering Eqs.~(\ref{eqn:diracexp}) 
and (\ref{eqn:kernelexp}), we find
\beqarr\label{eqn:ourbispectrum}
\Bis & = & \left(\frac{8}{\pi}\right)^2 \sum_{L \, \ell_1^\prime \ell_2^\prime}  
	   \!\! \frac{(i)^{\ell_1^\prime + \ell_2^\prime -\ell_1 - \ell_2}}{(2L+1)}
           \int \! dr \,r^2 \los \nonumber \\
     &   & \times \sum_{M m_1^\prime m_2^\prime} \, (-1)^{m_1 + m_2^\prime} \Gaunttwo \Gauntstarone 
\Gauntstartwo  \nonumber \\
     &	 & + 
	   \left(\begin{array}{c} 
           \ell_3 \leftrightarrow \ell_1 \\
           \ell_2 \leftrightarrow \ell_3 
           \end{array}\right)
           +
           \left(\begin{array}{c} 
           \ell_3 \leftrightarrow \ell_2 \\
           \ell_1 \leftrightarrow \ell_3 
           \end{array}\right) \; ,    	
\eeqarr
where we have used orthonormality of
%relations between 
spherical harmonics, and we have defined 
\beqarr
%\mathcal{L}_{\ell_1 \ell_2 \ell_3}^{\ell_1^\prime \ell_2^\prime \ell_3^\prime}(r) & 
%\equiv & \int dk_1 k_1^2 
%                         %\tilde{\Delta}_{\ell_1}(k_1) 
%                         \Delta_{\ell_1}(k_1) j_{\ell_1}(k_1 r) 
%\int dk_2 k_2^2  P_{\Phi}(k_2)
%			 \Delta_{\ell_2}(k_2) j_{\ell_2^\prime}(k_2 r) \nonumber  \\
%                     & \times & \int dk_3 k_3^2 P_{\Phi}(k_3) 
%\Delta_{\ell_3}(k_3) j_{\ell_3^\prime}(k_3 r) 
% 			 f_{\ell_1^\prime}(k_1,k_2,k_3) \label{eqn:losdefinition}  
\mathcal{L}_{L \ell_1 \ell_2}^{\ell_3 \ell_1^\prime \ell_2^\prime}(r) & 
\equiv & \int dk_3 k_3^2 
                         %\tilde{\Delta}_{\ell_1}(k_1) 
                         \Delta_{\ell_3}(k_3) j_{\ell_3}(k_3 r) 
\int dk_1 k_1^2  P_{\Phi}(k_1)
			 \Delta_{\ell_1}(k_1) j_{\ell_1^\prime}(k_1 r) \nonumber  \\
                     & \times & \int dk_2 k_2^2 P_{\Phi}(k_2) 
\Delta_{\ell_2}(k_2) j_{\ell_2^\prime}(k_2 r) 
 			 f_{L}(k_1,k_2,k_3) \label{eqn:losdefinition}  
\; .
\eeqarr
Formula (\ref{eqn:ourbispectrum}) is what we have been looking for: it describes 
the angular CMB bispectrum arising from the primordial potential [Eq.~(\ref{eqn:NGpotential})]. 
%Nevertheless, it should be stressed that the observable quantity is 
The angle-averaged bispectrum, $\Avbis$, 
%which 
is related to $\Bis$ by Eq.~(\ref{eqn:angavbis}), and an explicit 
expression for the angle-averaged bispectrum can be easily 
derived from %with some further manipulations of 
Eq.~(\ref{eqn:ourbispectrum}). We use the following relation of the Wigner symbols: %In particular, we can write
\beqarr\label{eqn:bis2avbis} 
\sum_{M m_1^\prime m_2^\prime} (-1)^{m_1+m_2^\prime} \Gaunttwo \Gauntstarone \Gauntstartwo & = &  
(-1)^{\ell_3 + L} \,  I_{\ell_1^\prime \, \ell_2^\prime \, \ell_3} \, I_{\ell_2^\prime \, 
\ell_2 \, L} \,
I_{\ell_1 \, \ell_1^\prime \, L} \left\{\begin{array}{ccc}
                  	        \ell_1 & \ell_2 & \ell_3 \\
                  	        \ell_2^\prime & \ell_1^\prime & L
               	             \end{array}\right\} \nonumber \\
& & \times \left(\begin{array}{ccc}
	            \ell_1 & \ell_2 & \ell_3 \\
		    m_1    & m_2    & m_3
                 \end{array}\right) \; ,
\eeqarr
%Here we have introduced the Wigner $6j$ symbols 
where $\left\{\begin{array}{ccc} \ell_1 & \ell_2 & \ell_3 \\ \ell_2^\prime & \ell_1^\prime & 
L \end{array}\right\}$ is the Wigner $6j$ symbol, 
and we have defined the quantities
\beq
I_{\ell_1 \, \ell_2 \, \ell_3} \equiv 
\sqrt{\frac{(2\ell_1+1)(2\ell_2+1)(2\ell_3+1)}{4\pi}}
				      \left(\begin{array}{ccc}	
					 \ell_1 & \ell_2 & \ell_3 \\
					  0      & 0      & 0	
				      \end{array}\right) \; .
\eeq
%Remembering Eq. (\ref{eqn:angavbis}) it is now immediate to see, 
%from the formulae (\ref{eqn:ourbispectrum}) and (\ref{eqn:bis2avbis}) that
Using these quantities, we obtain the final analytic formula of the angle-averaged
bispectrum with arbitrary kernels:
\beqarr\label{eqn:ouravbis}
\Avbis & = & \left(\frac{8}{\pi}\right)^2  \sum_{L=0}^N 
             \sum_{\ell_1^\prime \ell_2^\prime = 0}^\infty  
	     \!\! \frac{(i)^{\ell_1^\prime + \ell_2^\prime -\ell_1 - 
\ell_2}(-1)^{\ell_3+L}}
	     {2L+1} 
	     \,I_{\ell_1^\prime \, \ell_2^\prime \, \ell_3} \, I_{\ell_2^\prime \, 
\ell_2 \, L} \, 
	     \,I_{\ell_1 \, \ell_1^\prime \, L} \, \left\{\begin{array}{ccc}
                  	                            \ell_1 & \ell_2 & 
\ell_3 \\   \ell_2^\prime & \ell_1^\prime & L
               	                                \end{array}\right\} 
 \nonumber \\
       &   & \times \int \! dr \, r^2 \los +	
 		\left(\begin{array}{c} 
           	      \ell_3 \leftrightarrow \ell_1 \\
                      \ell_2 \leftrightarrow \ell_3 
                      \end{array}\right)
                          +
                \left(\begin{array}{c} 
                      \ell_3 \leftrightarrow \ell_2 \\
                      \ell_1 \leftrightarrow \ell_3 
                \end{array}\right) \; .
\eeqarr
We use this general relation to calculate numerically the CMB  
angle-averaged bispectrum for the class of inflationary 
models that produce potentials %like the one 
in the form of Eq.~(\ref{eqn:NGpotential}).
To select and study a specific model we need to provide an  
explicit expression for the coefficients of the Legendre 
expansion of the kernel, Eq. (\ref{eqn:kernelexpansion}) ({\it i.e.} 
we need to provide an explicit 
expression for $f_{L}(k_1,k_2,k_3)$ in Eq.~(\ref{eqn:losdefinition}) ). 
We will now consider the various possibilities for the kernels in the next subsections. %turn.

\subsection{Constant Kernel}
\noindent
The simplest possible choice of the kernel is a constant, $\kernel = f_{\rm NL}$ 
(where $f_{\rm NL}$ is a constant parameterizing the level of non-Gaussianity), which gives %corresponds to
\beq
\fix = \fiLx + f_{\rm NL} \left[\Phi_{\rm L}^2(\mathbf{x}) - 
\left\langle \Phi_{\rm L}^2(\mathbf{x}) \right\rangle \right] \; ,
\eeq
in real space. This is the usual phenomenological 
parametrization of non-Gaussianity which has been widely used in the literature. %so far.
The CMB angular bispectrum in this model %arising from this kind of primordial potential 
has been calculated by Komatsu and Spergel \cite{ks}. 
As a simple check of our calculations, we %want now to 
rederive their formula %result,
starting from Eq.~(\ref{eqn:ouravbis}). 

%The case we are exploring is obtained by substituting 
For a constant $f_{\rm NL}$, $N = 0$ and
$f_0(k_1,k_2,k_3) = f_{\rm NL}$ in Eq. (\ref{eqn:kernelexp}); thus,  
Eq.~(\ref{eqn:ouravbis}) yields
\beqarr\label{eqn:weakcoupling}
\Avbis & = & \left(\frac{8}{\pi}\right)^2 
             \sum_{\ell_1^\prime \ell_2^\prime = 0}^\infty  
	     \!\! (i)^{\ell_1^\prime + \ell_2^\prime -\ell_1 - \ell_2} (-1)^{\ell_3} 
	     \,I_{\ell_1^\prime \, \ell_2^\prime \, \ell_3} \, I_{\ell_2^\prime \, \ell_2 \, 0} \, 
	     \,I_{\ell_1 \, \ell_1^\prime \, 0} \, \left\{\begin{array}{ccc}
                  	                            \ell_1 & \ell_2 & \ell_3 \\
                  	                            \ell_2^\prime & \ell_1^\prime & 0
               	                                \end{array}\right\} \
 \nonumber \\
       &   & \times \int \! dr \, r^2 \loszero +	
 		\left(\begin{array}{c} 
           	      \ell_3 \leftrightarrow \ell_1 \\
                      \ell_2 \leftrightarrow \ell_3 
                      \end{array}\right)
                          +
                \left(\begin{array}{c} 
                      \ell_3 \leftrightarrow \ell_2 \\
                      \ell_1 \leftrightarrow \ell_3 
                \end{array}\right) \; .
\eeqarr
We can write
\beq
I_{\ell_2^\prime \ell_2 0} =  \sqrt{\frac{(2\ell_2+1)}{4 \pi}}
                       (-1)^{\ell_2} \delta_{\ell_2^\prime}^{\ell_2} \; ,
\eeq
where $\delta_{\ell_2^\prime}^{\ell_2}$ is a Kronecker delta 
and we have used the formula:
\beq                                                              
 \left(\begin{array}{ccc}
    \ell_1 & \ell_2 & 0 \\
       m   &   -m   & 0
 \end{array}\right) \nonumber \\ = (-1)^m 
\frac{(-1)^{\ell_1}}{\sqrt{2\ell_1+1}} \delta_{\ell_1}^{\ell_2} \, .
\eeq
An analogous relation holds for $I_{\ell_1 \ell_1^\prime 0}$, giving
\beqarr%\label{eqn:weakcoupling}
\Avbis & = & \left(\frac{8}{\pi}\right)^2  (-1)^{\ell_1+\ell_2 + \ell_3} 
	     \,\frac{(2\ell_3+1)^{\frac{1}{2}}(2\ell_2+1)
(2\ell_1+1)}{(4\pi)^{\frac{3}{2}}} \nonumber \\ 
       &   & \times \left\{\begin{array}{ccc}
                \ell_1 & \ell_2 & \ell_3 \\
                \ell_2 & \ell_1 & 0
             \end{array}\right\} 
             \left(\begin{array}{ccc}
		\ell_1 & \ell_2 & \ell_3 \\
                0      & 0      & 0
             \end{array}\right) \nonumber \\
       &   & \times \int \! dr \, r^2 \losfNL +	
 		\left(\begin{array}{c} 
           	      \ell_3 \leftrightarrow \ell_1 \\
                      \ell_2 \leftrightarrow \ell_3 
                      \end{array}\right)
                          +
                \left(\begin{array}{c} 
                      \ell_3 \leftrightarrow \ell_2 \\
                      \ell_1 \leftrightarrow \ell_3 
                \end{array}\right) \; .
\eeqarr
If we now make use of the relation
\beq
\left\{\begin{array}{ccc}
  \ell_1 & \ell_2 & \ell_3 \\
  \ell_2 & \ell_1 & 0
\end{array}\right\} = \frac{(-1)^{\ell_1+\ell_2+\ell_3}}
{\sqrt{(2\ell_2+1)(2\ell_1+1)}} \; ,
\eeq
and remember that, when $\kernel = f_{\rm NL}$, by definition 
(see Eq. (\ref{eqn:losdefinition})),
\beqarr
\losfNL & = & f_{\rm NL} \int dk_1 k_1^2 \Delta_{\ell_1}(k_1) 
P_{\Phi}(k_1) j_{\ell_1}(k_1 r) \int dk_2 k_2^2  
                         P_{\Phi}(k_2) \Delta_{\ell_2}(k_2) 
j_{\ell_2}(k_2 r) \nonumber \\
        &\times & \int dk_3 k_3^2 \Delta_{\ell_3}(k_3) j_{\ell_3}(k_3 r) \; ,
\eeqarr
we %immediately 
recover exactly the result of Ref. \cite{ks}.

\subsection{Momentum-dependent kernels}
\noindent 
Throughout the rest of this work we are going to
consider a primordial potential kernel defined
by Eq. (\ref{eqn:f_NL}):

\beq\label{eqn:ourkernel}
\kernel = -\frac{1}{6}-\frac{5}{3}\left(1-a_{\rm NL}\right) 
-  k_1 k_2
\frac{\angk_1 \cdot \angk_2}{k_3^2} + 3 \frac{k_1^2 k_2^2}{k_3^4} 
         +3k_1^2 k_2^2 \frac{(\angk_1 \cdot \angk_2)^2}{k_3^4} + 
3k_1 k_2 \frac{(k_1^2+k_2^2)(\angk_1 \cdot \angk_2)}{k_3^4}   \; ,
\eeq
%It is immediate to see that formula (\ref{eqn:ourkernel}) 
%is definition (\ref{eqn:kernelexpression}) with $N=2$.
It follows from the form of the kernel %quantity $k_3^{2N}\kernel = k_3^4\kernel$ 
that we can expand $\kernel$ into %terms of 
the first three Legendre polynomials in terms of the angle between 
$\mathbf{k_1}$ and $\mathbf{k_2}$:

\beqarr
%k_3^4 
\kernel                & = &  \sum_{\ell=0}^{2} f_\ell(k_1,k_2,k_3) 
                                P_\ell(\angk_1 \cdot \angk_2)  \; ,   
       \\
P_0(\angk_1 \cdot \angk_2) & = & 1                             
\; ,                \\
P_1(\angk_1 \cdot \angk_2) & = & \angk_1 \cdot \angk_2         
\; ,                \\
P_2(\angk_1 \cdot \angk_2) & = & \frac{1}{2}\left[ 3(\angk_1 
\cdot \angk_2)^2-1 \right] \; .
\eeqarr
A simple, direct calculation shows, for our kernel, that 
\beqarr
f_0(k_1,k_2,k_3) & = & \left[-\frac{1}{6}-\frac{5}{3}\left(1-a_{\rm NL}\right)
\right]%k_3^4 
+\frac{4 k_1^2 k_2^2}{k_3^4} \label{eqn:f0} \; , \\
               &   &           \nonumber       \\
f_1(k_1,k_2,k_3) & = & \frac{k_1 k_2}{k_3^2}\left[3\left(\frac{k_1^2+k_2^2}{k_3^2}\right)-%k_3^2
1\right]   
\; , \\
               &   &           \nonumber       \\
f_2(k_1,k_2,k_3) & = & \frac{2 k_1^2 k_2^2}{k_3^4} \label{eqn:f2}     \; .
\eeqarr
Therefore, we find that the conventional momentum-independent parameterization, $f_{\rm NL}$, 
captures only the first term in $f_0$.
%Our purpose is now to 
We evaluate numerically the expression of the CMB 
angle-averaged bispectrum, which is obtained 
by %after 
substituting these $f_\ell(k_1,k_2,k_3)$ coefficients 
%these last formulae 
into 
%Eqs.~(\ref{eqn:ouravbis})
%and 
(\ref{eqn:losdefinition}):

%Let us now focus our attention on the generation of $\los$.
%In our case we can obtain $\los$, for different values of $\ell$, as

\beqarr
\loszero & = & \left[-\frac{1}{6}-\frac{5}{3}\left(1-a_{\rm NL}\right)
               \right] \alpha_{\ell_3}^{(0)}(r)\beta_{\ell_1 \ell_1^\prime}^{(0)}(r)
	       \beta_{\ell_2 \ell_2^\prime}^{(0)}(r) + 4 \alpha_{\ell_3}^{(-4)}(r) 
\label{eqn:loszero}
	       \beta_{\ell_1 \ell_1^\prime}^{(2)}(r)\beta_{\ell_2 \ell_2^\prime}^{(2)}(r) 
\; , \\
	 &   &                                                     \nonumber 
\\   
\losone  & = & -\alpha_{\ell_3}^{(-2)}(r)\beta_{\ell_1 \ell_1^\prime}^{(1)}(r)
	       \beta_{\ell_2 \ell_2^\prime}^{(1)}(r) +
                3 \alpha_{\ell_3}^{(-4)}(r)\beta_{\ell_1 \ell_1^\prime}^{(3)}(r)
	       \beta_{\ell_2 \ell_2^\prime}^{(1)}(r) \nonumber \\                 
	 &   &  + 3 \alpha_{\ell_3}^{(-4)}(r)\beta_{\ell_1 \ell_1^\prime}^{(1)}(r)
	       \beta_{\ell_2 \ell_2^\prime}^{(3)}(r)                       \; , 
\label{eqn:losone} \\
	 &   &                            \nonumber             \\    
\lostwo  & = & 2 \alpha_{\ell_3}^{(-4)}(r)\beta_{\ell_1 \ell_1^\prime}^{(2)}(r)
	       \beta_{\ell_2 \ell_2^\prime}^{(2)}(r) \; ; \label{eqn:lostwo}
\eeqarr
the quantities $\alpha$ and $\beta$ being defined as
\beqarr\label{eqn:losfactor}
\alpha_{\ell}^{(n_1)}(r)       & \equiv & \int \! dk \, k^2 k^{n_1} 
\Delta_\ell(k) j_\ell(k r) 
				    \; ,\\
				  & & \nonumber \\
\beta_{\ell_1 \ell_2}^{(n_2)}(r) & \equiv & \int \! dk_1 \, 
k_1^2 k_1^{n_2} P_{\Phi}(k_1) \tfone 
                                      j_{\ell_2}(k_1 r) \; .
\label{eqn:beta}
\eeqarr
We then use these results in Eqs.~(\ref{eqn:ouravbis}) to compute the angle-averaged
bispectrum numerically.
%These results are obtained straightforwardly by substitution of the 
%Legendre expansion coefficients of $\kernel$, 
%formulae (\ref{eqn:f0}) to (\ref{eqn:f2}), into $\los$, defined by 
%Eq.~(\ref{eqn:losdefinition}).  

\section{Numerical results}\label{sec:results}

\subsection{Radial Coefficients}
\noindent
The problem of the numerical evaluation of $\Avbis$ can be divided
into two parts. The first is the calculation of
the Wigner $3j$ and $6j$ coefficients, 
while the other is the generation of the coefficients $\los$.
Since the expansion of our kernel contains only the first three 
Legendre polynomials, we consider %will be limited to the case 
only $0 \leq L \leq 2$.
This allows us to use analytic formulae %to compute 
of the $6j$ symbols. 
Also the $3j$ symbols in %appearing in the definitions of the 
%coefficients 
$I_{\ell_1 \ell_2 \ell_3}$ can be evaluated by %means of 
the well-known analytic formulae based on the Stirling approximation 
at high $\ell$'s.

The calculation of $\los$ can be reduced to the numerical evaluation 
of %the last two integrals
$\alpha_l^{(n_1)}(r)$ (Eq.~[\ref{eqn:losfactor}]) and $\beta_{l_1l_2}^{(n_2)}(r)$
(Eq.~[\ref{eqn:beta}]), in which
we have to account for all the possible choices of the set of values 
$\left\{L \, , \ell_1 \, , \ell_2 \, , \ell_3 \, , \ell_1^\prime \, , \ell_2^\prime 
\right\}$, while for $n_i$ we need only
$n_1=-4$, $-2$, and $0$, and $n_2=0$, 1, 2, and 3. (See Eq.~[\ref{eqn:loszero}-\ref{eqn:lostwo}]).
Eq.~(\ref{eqn:ouravbis}), applied to our case, shows that if we want to 
calculate a particular mode 
of the averaged bispectrum, $\Avbis$, we have to generate all the 
terms of $\los$ for $0 \leq L \leq 2$, 
$1 < \ell_1^\prime, \, \ell_2^\prime < \infty$. %In practice, due to 
The selection rules of the Wigner coefficients guarantee that 
the only terms which contribute to the sum over $\ell_1^\prime, \, \ell_2^\prime$ 
(for fixed $L$) are those of which satisfy the triangular conditions:
$\ell_1 - L \leq \ell_1^\prime \leq \ell_1 + L$ and $\ell_2 - 
L \leq \ell_2^\prime \leq \ell_2 + L$. 

\begin{figure} [h]
\begin{center}
\includegraphics[width = 0.8\textwidth]{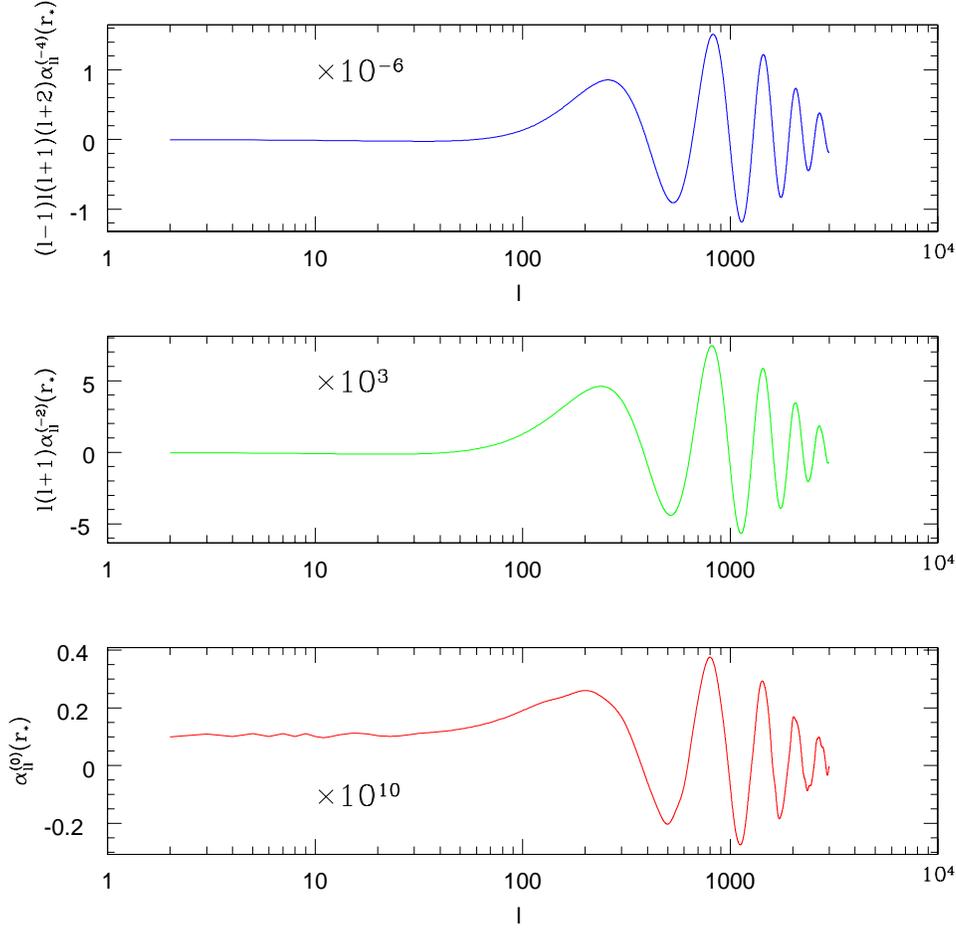}
\caption{Radial coefficients $\alpha^{(n)}_\ell(r)$ [Eq.~(\ref{eqn:losfactor})] at the time 
of decoupling, $r_*$. From top to bottom we plot  
$(\ell-1)\ell(\ell+1)(\ell+2)\alpha^{(-4)}_\ell(r_*)$, 
$\ell(\ell+1)\alpha^{(-2)}_\ell(r_*)$, $\alpha^{(0)}_\ell(r_*)$, respectively. 
%(see the text for more details). The cosmological model is a Concordance
%Model with $\Omega_{\Lambda} = 0.7$, $\Omega_b = 0.05$, 
%$\Omega_m = 0.3$, $h=0.65$ and $n=1$.
}
\label{fig:alphafig}
\end{center}
\end{figure}

\begin{figure} [h]
\begin{center}
\includegraphics[width = 0.8\textwidth]{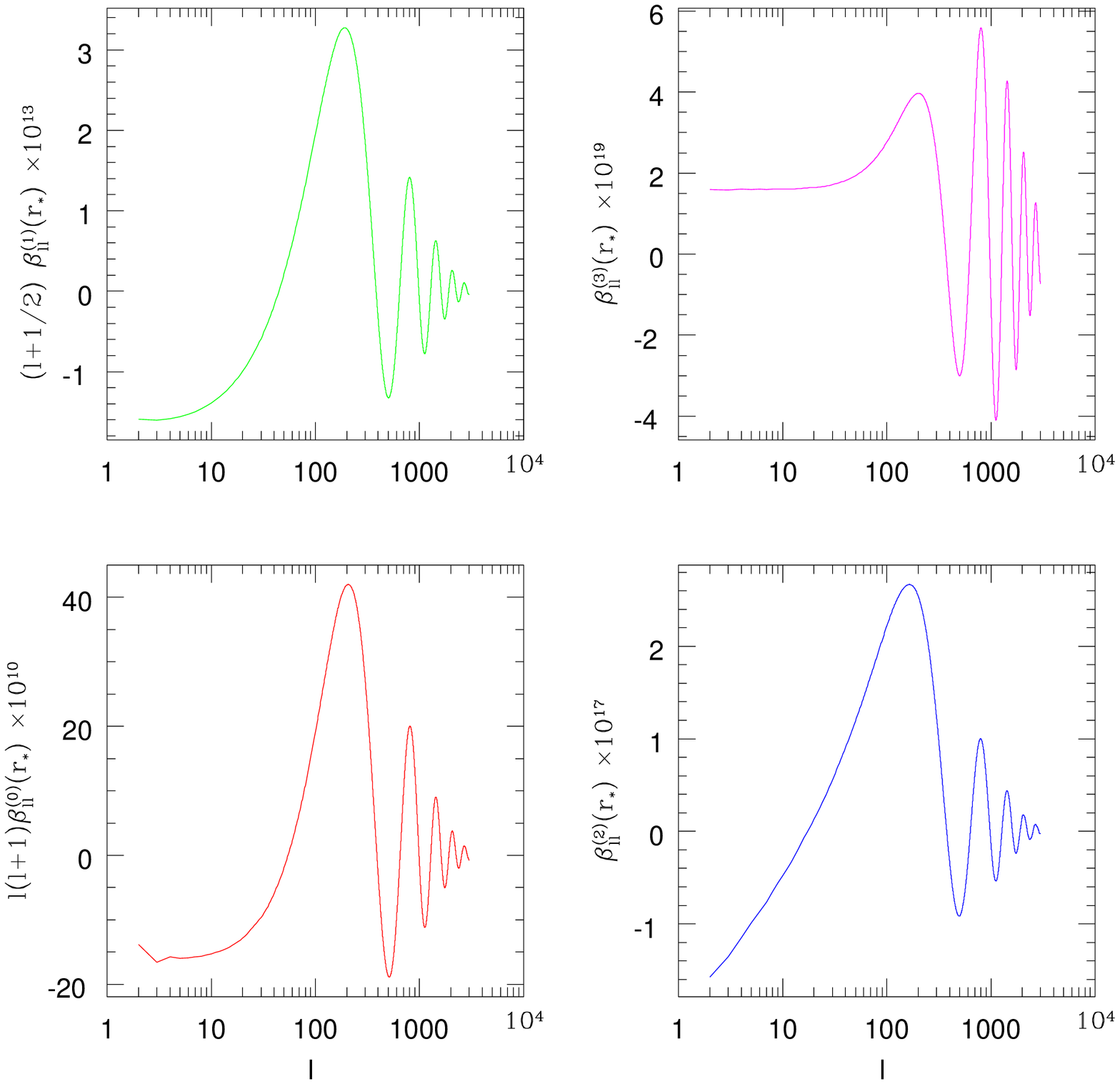}
\caption{Radial coefficients $\beta^{(n)}_l(r)$ [Eq.~(\ref{eqn:beta})] at the time 
of decoupling, $r_*$. On the left side, from top to bottom: 
$(\ell+ {1/2}) \beta^{(1)}_{\ell \ell}(r_*)$, $\ell(\ell+1)\beta^{(0)}_{\ell \ell}(r_*)$. 
On the right side, from top to bottom: $\beta^{(3)}_{\ell \ell}(r_*)$, 
$\beta_{\ell \ell}^{(2)}(r_*)$. 
%The cosmological model is a Concordance 
%Model with $\Omega_{\Lambda} = 0.7$, $\Omega_b = 0.05$, $\Omega_m = 0.3$, 
%$h=0.65$ and $n=1$. More details can be found in the text.
 }\label{fig:betafig}
\end{center}
\end{figure}

In our analysis, we consider a Concordance Model with 
$\Omega_{\Lambda} = 0.7$, $\Omega_b = 0.05$, $\Omega_m = 0.3$, 
$h=0.65$ and $n=1$;  
Figures (\ref{fig:alphafig}) and (\ref{fig:betafig}) show some 
radial coefficients $\alpha_\ell^{(n_1)}(r)$ 
and $\beta^{(n_2)}_{\ell \ell}(r)$, calculated at the 
time of decoupling, %$\tau_*$ (by definition 
$r_*=c(\tau_0 - \tau_*)$, where $\tau$ 
denotes conformal time, $\tau_0$ is the present conformal time, and the
decoupling time, $\tau_*$, is defined at the peak of visibility function. In our model we have $c\tau_0=14.9$ Gpc and 
%the decoupling epoch is at 
$c\tau_*=289$ Mpc. 
%and 
%For all the values of $n$ contained in the formulae (\ref{eqn:loszero}),
%(\ref{eqn:losone}), (\ref{eqn:lostwo}). 
To calculate the radial coefficients we use a modified version of the CMBfast 
code\footnote{In our numerical computation we are neglecting second order corrections to 
the CMB radiation transfer functions. These should in principle be included for a complete 
and definitive treatment of CMB non-Gaussianity.}. 

Although most of the signal is generated in a narrow region around 
decoupling ({\it i.e.} when $r \sim r_* = c(\tau_0 - \tau_*) \sim c\tau_0$), 
in the low-$\ell$ regime we still have to account for the 
low-$r$ contribution due to the late integrated Sachs-Wolfe effect. Thus our 
$r$-integration boundary is $c(\tau_0-6\tau_*) < r < c\tau_0$ 
for $\ell>50$, whereas $0 < r < c\tau_0$ for $\ell \leq 50$. 
The step-size $\Delta r$ is determined by the ratio of the 
width of the last scattering surface to the present cosmic
horizon, $c\tau_0$, and by the necessity of an accurate sampling of 
the acoustic oscillations at recombination. As the number
of oscillations increase at high-$\ell$, we need smaller and 
smaller step sizes when simulating experiments with higher and higher angular 
resolutions.

\subsection{Signal-to-noise ratio for WMAP}
\noindent
Even if a significant angular bispectrum was detected in CMB, 
this would not necessarily mean that it was generated
by some primordial mechanism like inflation. There are in fact several 
foregrounds which can produce non-Gaussianity in CMB anisotropies,
such as the Sunyaev-Zel'dovich (SZ) effect, weak lensing, the presence 
of point sources, and so on. Thus a complete study of %the
detectability of the primary bispectrum needs to include %in principle
%also the calculation of 
the secondary bispectra generated by the foregrounds, 
in order to check if the primordial component can be isolated from others.   

Having calculated numerically the angle-averaged bispectrum 
from primary and secondary sources, %following Ref.~ \cite{ks}
%it is possible to introduce 
we evaluate a $\chi^2$ statistic\cite{ks}

\beq
\chi^2  = \sum_{2 \leq \ell_1 \leq \ell_2 \leq \ell_3} 
\frac{\left(B_{\ell_1 \ell_2 \ell_3}^{obs} - 
        \sum_i A_i B_{\ell_1 \ell_2 \ell_3}^{(i)} 
\right)}{\sigma^2_{\ell_1 \ell_2 \ell_3}} \; ,
\eeq
where $B_{\ell_1 \ell_2 \ell_3}^{obs}$ is the observed bispectrum and 
$B_{\ell_1 \ell_2 \ell_3}^{(i)} $ are the 
theoretically calculated bispectra for different components, denoted by $i$. 
The variance $\sigma^2_{\ell_1 \ell_2 \ell_3}$ of the bispectrum
can be written as \cite{sg,mg}

\beq
\sigma^2_{\ell_1 \ell_2 \ell_3} = \left\langle B_{\ell_1 \ell_2 
\ell_3}^2\right\rangle -
                                \left\langle B_{\ell_1 \ell_2 
\ell_3}\right\rangle^2 \simeq C_{\ell_1} C_{\ell_2} C_{\ell_3} 
				\Delta_{\ell_1 \ell_2 \ell_3} \; ,
\eeq
where $\Delta_{\ell_1 \ell_2 \ell_3}$ takes values $1$, $2$, and $6$ 
when all $\ell$'s are different, two of them are equal and all 
are the same, respectively. $C_\ell$ is the sum of the theoretical 
CMB angular power-spectrum and the power-spectrum of the detector 
noise. The last one can be calculated analytically using Ref.~\cite{knox}.

Taking ${\partial \chi^2 / \partial A_i} = 0$, the Fisher matrix %can be
%defined as \cite{ks}
is given by \cite{ks}

\beq\label{eqn:fisher}
F_{ij} = \sum_{2 \leq \ell_1 \leq \ell_2 \leq \ell_3} 
\frac{ B_{\ell_1 \ell_2 \ell_3}^{(i)} B_{\ell_1 \ell_2 \ell_3}^{(j)}}{
              \sigma^2_{\ell_1 \ell_2 \ell_3}} \; ,
\eeq
and the signal-to-noise ratio $({S/N})_i$ for a component $i$ is %then defined as

\beq\label{eqn:SovN}
\left(\frac{S}{N}\right)_i = \frac{1}{\sqrt{F_{ii}^{-1}}} \; .
\eeq
Let us neglect for the moment the non-diagonal components of the 
Fisher matrix; then, denoting the primordial component by $i=1$, we can 
give an estimate of the expected signal-to-noise ratio for the 
primordial non-Gaussian 
signal without considering foregrounds. It is simply

\beq\label{eqn:SovNapprox}
\left(\frac{S}{N}\right)_1 \sim \sqrt{F_{11}} \; .
\eeq
We have calculated the approximated signal-to-noise ratio using formula 
(\ref{eqn:SovNapprox}) for an experiment with 
the FWHM beam-size of {\sl WMAP} ($FWHM = 13^{'}$, $\ell_\textrm{max} = 500$), 
assuming different scenarios for the generation of the cosmological
perturbations, namely the standard single-field slow roll scenario, the 
inhomogeneous reheating scenario and the curvaton scenario.  
The shape of the kernel in all these scenarios is given by %the one we assumed 
%in 
Eq. (\ref{eqn:ourkernel}) with model-dependent values of the constant 
$a_{\rm NL}$. According to Ref.~\cite{komreview}, in single-field slow 
roll inflation $a_{\rm NL} = 1$, in the 
inhomogeneous reheating case $a_{\rm NL} = {1/4}$, whereas in the curvaton scenario 
we have $a_{\rm NL} = ({3/4r}) - {r/2}$, where $r$ is the relative curvaton 
contribution to the total energy density at curvaton decay.

Let us now comment on our results, starting from the standard single-field inflation
and the inhomogeneous reheating cases.
%Even in the simplified approach we are considering ({\it i.e.} 
%neglecting foregrounds), 
Even though we ignore correlations between the primordial and secondary bispectra,
we find, for the standard inflationary scenario,
 the expected signal-to-noise ratio for {\sl WMAP} is ${S/N} \simeq 0.10$ and, for the 
 inhomogeneous reheating case, %a signal-to-noise ratio 
 ${S/N} \simeq 0.15$; thus,
%meaning that 
the primordial non-Gaussianity from these models is %clearly 
below the {\sl WMAP} detection 
threshold. As the correlation between the primordial and secondary bispectra would only lower 
the signal-to-noise ratio for the primordial component, we can conclude that 
the primordial bispectrum from these scenarios
%standard inflation 
is undetectable with {\sl WMAP}.%present experiments, 
%even before accounting for secondary sources.

These expectations confirm the ones obtained in the previous work %phenomenological
%approach to primordial non-Gaussianity, 
where 
the non-Gaussian primordial gravitational potential was approximated as 
$\Phi_{\rm NL} = \Phi_{\rm L} + f_{\rm NL} \Phi_{\rm L}^2$
and $f_{\rm NL}$ is a %constant 
momentum-independent 
parameter defining the level of predicted non-Gaussianity.
In this framework Komatsu and Spergel 
derived a detection threshold of $f_{\rm NL} = 20$ for {\sl WMAP} and
$f_{\rm NL} = 5$ for {\sl Planck}. This suggested that a primordial signal
from standard single-field inflation would be undetectable by {\sl WMAP}
as, in this phenomenological approach, $f_{\rm NL}$ was expected to be $\simeq 1$
in the standard scenario \cite{ks, Gangui, Salopek, Carroll}. 

\subsection{Challenges of numerical calculations at high $\ell$}
\label{sec:numerical}
%Before concluding this section, 
Before we study the expected signal-to-noise ratio for the future high-resolution
experiments such as {\sl Planck}, 
we %want to stress 
note that the kind 
of computation we have described so far is numerically very challenging. 
Even after parallelizing and optimizing as much as possible, our algorithm
({\it e.g.} by implementing analytic approximations for the Wigner
coefficients and by minimizing the number of points in the integration 
samples), the highest $\ell_\textrm{max}$ we can reach is $\ell_\textrm{max} = 500$, 
corresponding to the angular resolution of the {\sl WMAP} 
satellite. 
% As we explained in the previous section, 
We have not been able to go beyond the {\sl WMAP} resolution, as 
the CPU time requirement was too demanding. The parallelized version of our algorithm took
$6$ hours on $100$ processors to calculate the full bispectrum up to
$\ell_\textrm{max}=500$. As the CPU time scales roughly as $\ell_\textrm{max}^5$ 
it would take about 5 years on the same number of
processors to calculate it up to $\ell_\textrm{max}=3000$, thus making
approximations necessary.
On the other hand, extrapolating our results 
to higher angular resolutions suggests that the primordial non-Gaussian signal could 
be significant enough to allow detection of the primary signal 
at $\ell_{max} = 3000$ (the angular resolution of the {\sl Planck} satellite),
even in the most standard single-field slow roll inflationary scenario. 
This will be explained in more detail in the next section.       

\subsection{Prospects for detecting non-Gaussianity by {\sl Planck}}

Komatsu and Spergel \cite{ks} %also
pointed out that even an ideal experiment needed $f_{\rm NL} > 3$ in order
to detect primordial non-Gaussianity. This last statement, when combined with 
the previous theoretical expectations of the amplitude of non-Gaussianity,
$f_{\rm NL} \simeq 1$, implied rather 
pessimistic prospects for detecting the primordial 
non-Gaussian signals in standard scenarios of single-field inflation.

However, we stress here that the previous expectation, $f_{\rm NL} \simeq 1$,
even though it roughly took into account the effect of the post-inflationary evolution of non-Gaussianity,
was not based on the detailed second-order computation of the cosmological perturbations during and 
after inflation. For this reason, it must be considered only as an order of 
magnitude estimate, and care must be taken when we study detectability of
%which does not allow to say conclusively whether or not primordial 
non-Gaussianity from standard single-field inflation %can be detected 
by the future experiments at high angular resolution such as {\sl Planck}.  
 
%In fact 
Our prediction %we presented here, 
based on the complete second-order 
%Rapproach to the 
calculation of the primordial gravitational 
potential shows an encouraging trend which shows that the actual signal-to-noise ratio
is larger than the previous prediction with $f_{\rm NL}=1$. %an enhancement of the expected primordial non-Gaussian signal 
%with respect to this previous estimate. This enhancement, 
Even though it is insufficient to push the primary signal over the detectability threshold of
{\sl WMAP}, it could be big enough to allow detection of the primordial
non-Gaussianity signals by %the 
%forecoming 
{\sl Planck}. %experiment even in the standard 
%single-field scenario. 

\begin{figure} [!h]
\begin{center}
\includegraphics[width = 0.8\textwidth]{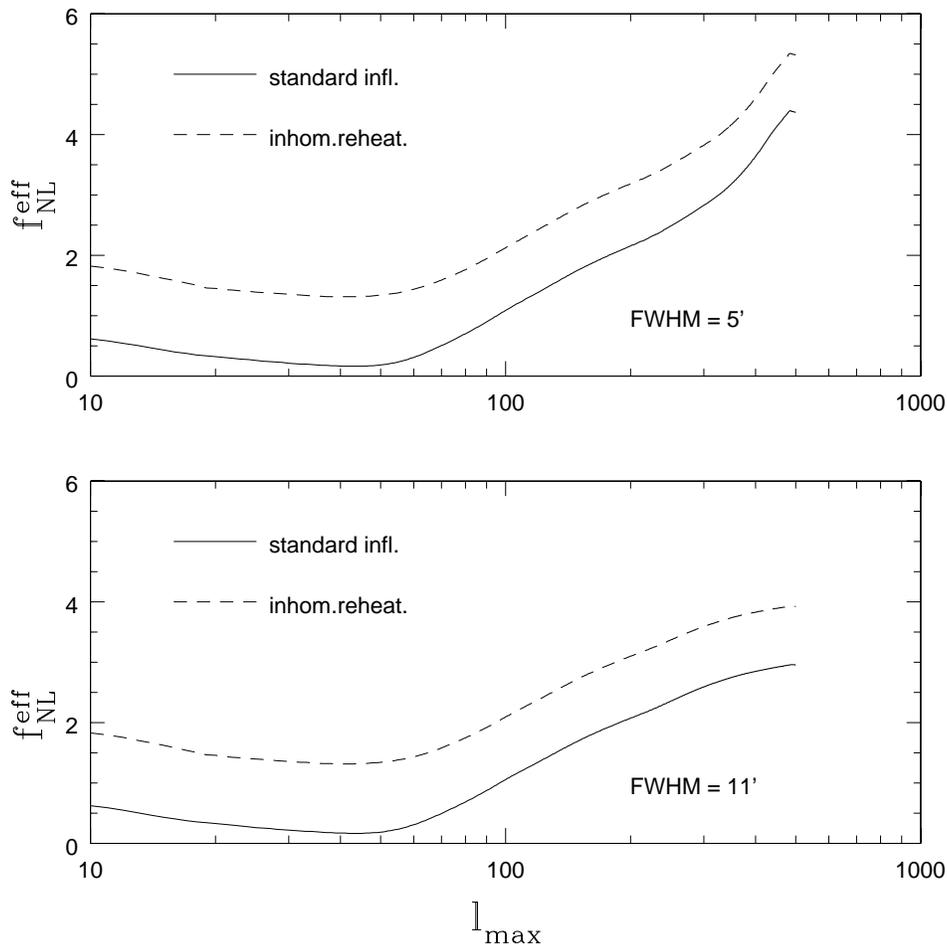}
\caption{Values of $f_{\rm NL}^{eff}(\ell_{\rm max})$ (Eq.~[\ref{eqn:fnleffdef}]) in the standard single-field and 
inhomogeneous reheating scenarios. 
This parameter, $f^{eff}_{\rm NL}(\ell_{\rm max})$, represents $f_{\rm NL}$ in the usual parametrization 
of non-Gaussianity to reproduce 
the same level of non-Gaussianity predicted by our model for a given $\ell_{\rm max}$.
For the standard single-field inflation 
the contribution to non-Gaussianity comes {\em only} from the post-inflation 
non-linear processing of perturbations, which is independent of the inflationary model. 
Thus the solid line in the plots also represents that part 
of the non-Gaussian signal which {\em must} be present in the CMB anisotropies, regardless of 
the considered inflationary model. The lower panel shows our results 
for an experiment with beam size and noise characteristics similar to {\sl WMAP}.
The upper panel shows the same analysis for {\sl Planck}. 
We are considering $\ell_\textrm{max} = 500$, corresponding
to the angular resolution of {\sl WMAP}. A full analysis for {\sl Planck} would require 
$\ell_\textrm{max} = 3000$, which is beyond the current computational power
(see Sec~\ref{sec:numerical}).
%possibilities of our 
%algorithm. Nonetheless we notice a clear enhancement of the signal with 
%respect to previous theoretical predictions already at this stage
% (previous order of magnitude estimates for standard single-field
%inflation gave $f_{\rm NL} \simeq 1$).
}\label{fig:fNLeff}
\end{center}
\end{figure}

Let us elaborate on this point. We evaluate %started by evaluating the 
the signal-to-noise ratio for %corresponding to different values of 
$10<\ell_\textrm{max}<500$ both in our cases and in the standard $f_{\rm NL}$ 
parametrization. 
In the standard parametrization of non-Gaussianity the 
signal-to-noise ratio can be written as\footnote{%Remember that 
We ignore %are neglecting 
the contributions 
%to non-Gaussianity coming 
from the foregrounds.}:
\beq
\left(\frac{S}{N}\right)^{\rm stand} = f_{\rm NL} \sqrt{F^{\rm stand}_{11}(f_{\rm NL}=1)} \; ,
\eeq
where $F^{\rm stand}_{11}(f_{\rm NL}=1)$ is %indicates the value of the component $11$ of 
the Fisher matrix of the standard $f_{\rm NL}$ model with %computed using formula (\ref{eqn:fisher}) and taking 
$f_{\rm NL}=1$. The idea is as follows: by comparing the actual signal-to-noise ratio
predicted from our full calculations, $(S/N)^{\rm full}$, to the standard one, 
we can estimate $f_{\rm NL}$ that is required to  produce the same $(S/N)^{\rm stand}$ 
in the standard parametrization as $(S/N)^{\rm full}$:
% in order to produce the same level of $(S/N)^{\rm full}$.
%the expected signal-to-noise ratio:
\beq\label{eqn:fnleffdef}
f^{eff}_{\rm NL}(\ell_{\rm max}) \equiv \frac{ \left({S/N}\right)^{\rm full} }{ \sqrt{F^{\rm stand}_{11}(f_{\rm NL}=1)}}
\; ,
\eeq
%where $({S/N})^{\rm new}$ is our new computation of the signal-to-noise ratio.
%This $f_{\rm NL}$ can be regarded as a sort of ``effective'' 
%$f_{\rm NL}$  for our model, meaning that 
which is the $f_{\rm NL}$ 
that is needed in the usual parametrization of non-Gaussianity to reproduce 
the same level of non-Gaussianity predicted by our model for a given $\ell_{\rm max}$. %For 
%this reason, from now on we will indicate it using the symbol 
%$ f_{\rm NL}^{\rm eff}$. The introduction of 
This parameter allows us to compare 
the previous estimates to ours more easily.
The results are shown in figure \ref{fig:fNLeff}, where we consider 
two experiments with the beam and the noise characteristics similar to {\sl WMAP} and 
{\em{Planck}}. Two 
things are worth noticing. First of all, %we see that our 
$f_{\rm NL}^{eff}(\ell_{\rm max})$ is not constant over $\ell_{\rm max}$. Second of all, %the value 
%of 
$f_{\rm NL}^{eff}(\ell_{\rm max}=500)$ %that we obtain when $\ell = 500$  
is significantly bigger than 
the previously expected value, $f_{\rm NL} \simeq 1$, though it is still of the same 
order of magnitude. 
When we look at the {\sl Planck} experiment, we also notice  
that $f_{\rm NL}^{eff}(\ell_{\rm max})$ is monotonically increasing when $\ell_{\rm max} \geq 40$, reaching a value of
$f_{\rm NL}^{eff}(\ell_{\rm max}=500) \simeq 4$,% for $\ell = 500$, which
which is already very close to the detection threshold $f_{\rm NL} = 5$ computed by 
Komatsu and Spergel \cite{ks} for the full resolution of {\sl Planck}. Considering that the angular
resolution of the {\sl Planck} satellite corresponds to $\ell_\textrm{max}=3000$, and 
we stopped our computation at $\ell_\textrm{max} = 500$, our results suggest that the 
non-Gaussian signal from standard single-field inflation is likely to be detected by {\sl Planck}.
%the {\sl Planck} satellite.

In addition to
%Besides considering 
the standard single-field inflation and the inhomogeneous
 reheating models, we also investigate the curvaton scenario.
In this case the value of $a_{\rm NL}$ depends on the parameter $r$, the relative curvaton 
contribution to the total energy density at curvaton decay, as previously
pointed out. We consider different values of $r$, and %, even in this case,
%we compared our predicted value of the signal-to-noise ratio with the 
%old-predicted one, which is obtained assuming the standard $f_{\rm NL}$ parametrization
calculate $f_{\rm NL}^{\rm eff}(\ell_{\rm max})$.
Note that the momentum-independent part has been calculated as $f_{\rm NL} = -{5/(4r)} + {5r/6}$ \cite{BMR3}. 
Our results are summarized in 
table \ref{tab:curvaton}. We notice
that, for small values of $r$, the parameter $f_{\rm NL}^{eff}(\ell_{\rm max}=500)$ is now 
smaller than what was expected in the previous predictions (meaning that 
our signal-to-noise ratio is smaller than what was predicted assuming the standard $f_{\rm NL}$ 
parametrization), whereas $f_{\rm NL}^{eff}(\ell_{\rm max}=500)>f_{\rm NL}$ for 
$r \geq 0.5$. %,  we see an enhancement of the signal. Also in this case we conclude 
%that a full reanalysis of previous phenomenological predictions will be required at high
%angular resolutions.
Therefore, it is incorrect to conclude that the amplitude of non-Gaussianity is smaller for 
larger $r$; on the contrary, the signal-to-noise stays nearly the same regardless of $r$.

\begin{table}[!t]
\small
\begin{center}
\begin{tabular}{|c|c|c|c|}
\hline
$\;\;\;\; \mathbf{r} \;\;\;\;$    & $\;\;\;\;\; \mathbf{|f_{\rm NL}|} \;\;\;\;\;$    
& $  \;\;\;\;\; \mathbf{f_{\rm NL}^{eff}} \;\;\;\;\; $    & $ \;\;\;\;\; \mathbf{{S/N}} \;\;\;\;\; $ \\
\hline
$0.1$   &    $12.42$    &    $8.42$     &    $0.24$ \\
\hline
$0.2$   &    $6.10$     &    $2.98$     &    $0.08$ \\
\hline
$0.3$   &    $3.92$     &    $2.62$    &    $0.06$ \\
\hline
$0.4$   &    $2.79$     &    $2.58$    & $0.07$ \\
\hline
$0.5$   &    $2.10$     &    $2.98$    & $0.08$ \\
\hline
$0.6$   & $1.58$        & $3.31$       & $0.10$ \\
\hline
$0.7$ & $1.20$ & $3.60$ & $0.10$ \\
\hline
$0.8$ & $0.90$ & $3.84$ & $0.10$ \\  
\hline
$0.9$ & $0.64$ & $4.04$ & $0.11$ \\
\hline
$1.0$ & $0.42$ & $4.22$ & $0.12$ \\
\hline
\end{tabular}
\caption{\rm Results for the curvaton model. The first two columns show the
value of the relative curvaton contribution to the total energy density
at curvaton decay and the predicted values of $f_{\rm NL}$ in the previous
parameterization which assumes that $f_{\rm NL}$ is a constant.
 The last two columns contain the new computation of $f_{\rm NL}^{eff}(\ell_{\rm max}=500)$ 
 (Eq.~[\ref{eqn:fnleffdef}])
 and of
the signal-to-noise ratio for {\rm WMAP}.}%. We are considering an experiment with the beam size of WMAP.} 
\label{tab:curvaton} 
\end{center}   
\end{table}

Before concluding this section, let us stress again that our rough estimate 
of $f_{\rm NL}^{eff}(\ell_{\rm max})$ and the 
extrapolation of our results to the angular resolution of the 
{\sl Planck} satellite  
does not allow any conclusive statement about detectability of 
the primordial non-Gaussian signals generated by the simplest models of 
inflation. It is important to keep in mind that our $f_{\rm NL}^{eff}(\ell_{\rm max})$ depends on
$\ell_{\rm max}$; thus, it is still possible that $f_{\rm NL}^{eff}(\ell_{\rm max})$ 
might start to decrease for $\ell_{\rm max} > 500$ and stay below the detection threshold of 
{\sl Planck} at $\ell_{\rm max} = 3000$.
%On the other hand, concerning this issue, the monotonically increasing 
However, we find an encouraging trend that $f_{\rm NL}^{eff}(\ell_{\rm max})$ 
increases monotonically for $\ell_{\rm max} \gtrsim 40$.
%behaviour of $f_{\rm NL}^{eff}$ in the range $40 < \ell < 500$ looks very promising and 
%suggests that it is worth
It is certainly worth finding a way to achieve the full numerical computation of the 
primordial bispectrum (\ref{eqn:ouravbis}) at very high multipoles 
($\ell_\textrm{max} \simeq 3000$). The current algorithm %Unfortunately a direct approach, 
based on the full numerical integration of equation~(\ref{eqn:ouravbis})
is computationally very expensive, and %very
%challenging, even assuming the possibility of further improvements 
%of our algorithm. 
most likely some approximations must be invoked; %introduced to
%make an exact numerical analysis affordable; 
one possibility is to implement the flat sky approximation at high $\ell$'s. This will be 
the topic of a forthcoming publication \cite{flatsky}.

\section{Conclusions}\label{sec:conclusions}

%In this paper our starting point has been the 
%general expression for the second-order  gravitational potential produced
%in the various mechanisms for the generation of the 
%cosmological perturbations during or  after inflation. The main feature
%of the associated nonlinearity parameter is to present a 
%non-trivial momentum-dependent kernel. 
In this paper, we have shown that the full second-order calculations
of cosmological perturbations and inflationary dynamics suggest that
the realistic form of non-Gaussianity, the kernel $f_{\rm NL}({\mathbf k}_1,{\mathbf k}_2, \mathbf{k_3})$, 
must contain momentum-dependent terms.
We have derived the analytic formula  for the angle-averaged primary CMB angular bispectrum. 
This formula allows a more realistic  description of non-Gaussian CMB anisotropy, 
extending the phenomenological model adopted in Ref.~\cite{ks}, where $f_{\rm NL}$ was taken to be a constant. 
We have developed a numerical 
code to compute the primary bispectrum and estimated the expected 
signal-to-noise ratio
for detecting primary non-Gaussianity at the {\sl WMAP} angular resolution. Our
results show that, in the framework of standard single-field inflation, 
the primary non-Gaussian signal cannot be detected by {\sl WMAP}, as already 
indicated by the previous analysis. On the other hand, in our complete 
second-order approach to perturbations during and after inflation, 
we have found that the previous theoretical expectation, $f_{\rm NL}\simeq 1$, was
too pessimistic, and the actual value which defines the CMB bispectrum
is much larger. 
%an enhancement of the signal-to-noise ratio, when compared to 
%previous approximate theoretical estimates, which could make the signal 
This result implies that the primordial non-Gaussian signals might be 
detectable by the future {\sl Planck}  
mission even in the standard single-field scenarios of inflation. 
However, using the current numerical algorithm, we have not been able to
reach {\sl Planck}'s angular resolution, $\ell_{\rm max}=3000$,
which would require 5 years of CPU time on 100 processors,
and our conclusion on the prospect for detecting non-Gaussianity by {\sl Planck}
has to rely on extrapolations from $\ell_{\rm max}=500$.
%This last statement has been obtained only by roughly extrapolating 
%our results to higher angular resolutions but 
%suggests that a full numerical computation at very high multipoles 
%($\ell_\textrm{max} = 3000$) could give interesting results.
%Unfortunately, an exact computation of the bispectrum at such high 
%resolutions constitutes a big numerical challenge. 
Suitable approximations
at high $\ell$'s will be required in the future, in order to make a definitive
conclusion on detectability of primordial non-Gaussianity in CMB.
%handle the problem numerically.
Finally, let us comment on statistical methods to measure the bispectrum.
Komatsu, Spergel and Wandelt \cite{ksw} have shown that the direct measurement
of all possible configurations of the bispectrum is computationally
too expensive, and developed a faster estimator of $f_{\rm NL}$
assuming that $f_{\rm NL}$ is a constant. Recently, Creminelli et al.
\cite{paolo} have extended this method to the case where the dominant
signals come from the equilateral configurations, which
yields a certain momentum-dependence in $f_{\rm NL}$. 
Their model (Eq.~[14] of \cite{paolo}), however, is different
from the form of $\kernel$ in equation~(\ref{eqn:ourkernel}), and thus
their estimator cannot be used to measure primordial non-Gaussianity
from second-order perturbations. New estimators optimized to 
our $\kernel$ need to be developed.

\acknowledgments{}
F.H. was supported by a Marie Curie
European Reintegration Grant within the 6th European Community
Framework Programme.
E.K. acknowledges support from an Alfred P. Sloan Fellowship.

%%%%%%%%%%%%%%%%%%%%%%%%%%%%%%%%%%%%%%
%%%%%%%%%%%%%%%%%%%%%%%%%%%%%%%%%%%%%%

%%%%%%%%%%%%%%%%%%%%%%%%%%%%%%%%%%%%%%
%%%%%%%%%%%%%%%%%%%%%%%%%%%%%%%%%%%%%%

\begin{thebibliography}{99}
\frenchspacing
%%%%%%%%%%%%%%%%%%%%%%%%%%%%%%%%%%%
%%%%%%%%%%%%%%%%%%%%%%%%%%%%%%%%%%%
\bibitem{lrreview} D.~H.~Lyth and A.~Riotto, Phys.\ Rept.\ 314 (1999) 1.

\bibitem{bennett/etal:2003b}
C.~L.~Bennett, et al., Astrophys.\ J.\ Suppl.\ 148 (2003) 1.

\bibitem{ex} 
H.~V.~Peiris, et al.,
%``First year Wilkinson Microwave Anisotropy Probe ({\sl WMAP}) observations:  
%Implications for inflation,''
Astrophys.\ J.\ Suppl.\ 148 (2003) 213.

\bibitem{Mollerach}
S.~Mollerach,
%``Isocurvature Baryon Perturbations And Inflation,''
Phys.\ Rev.\ D 42 (1990) 313.

\bibitem{curvaton1} K.~Enqvist and M.~S.~Sloth,
%``Adiabatic CMB perturbations in pre big bang string cosmology,''
Nucl.\ Phys.\ B 626 (2002) 395.

\bibitem{LW2}
D.~H.~Lyth and D.~Wands,
%``Generating the curvature perturbation without an inflaton,''
Phys.\ Lett.\ B 524 (2002) 5.

\bibitem{curvaton3} T.~Moroi and T.~Takahashi,
%``Effects of cosmological moduli fields on cosmic microwave background,''
Phys.\ Lett.\ B 522 (2001) 215 
[Erratum-ibid.\ B 539 (2002) 303].

\bibitem{LUW}
D.~H.~Lyth, C.~Ungarelli and D.~Wands,  
Phys.\ Rev.\ D 67 (2003) 023503.

\bibitem{komreview} N. Bartolo, E. Komatsu, S. Matarrese and A. Riotto,
Phys. Rept. {\bf }  (2005).

\bibitem{gamma} G.~Dvali, A.~Gruzinov and M.~Zaldarriaga,
%``A new mechanism for generating density perturbations from inflation,''
Phys.\ Rev.\ D 69 (2004) 023505; L.~A.~Kofman, arXiv:astro-ph/0303614.

\bibitem{val} E.W. Kolb,  A. Riotto and A. Vallinotto, 
  %``Curvature perturbations from broken symmetries,''
  Phys.\ Rev.\ D {\bf 71} (2005) 043513.

\bibitem{acqua}
V.~Acquaviva, N.~Bartolo, S.~Matarrese and A.~Riotto,
Nucl.\ Phys.\ B 667 (2003) 119.

\bibitem{maldacena}
J.~Maldacena, JHEP\ 0305 (2003) 013.

\bibitem{BMR2}
N.~Bartolo, S.~Matarrese and A.~Riotto, JHEP\ 0404 (2004) 006.

\bibitem{BMR3}
N.~Bartolo, S.~Matarrese and A.~Riotto, 
Phys.\ Rev.\ D 69 (2004) 043503.

\bibitem{BMR4}
N.~Bartolo, S.~Matarrese and A.~Riotto, 
JCAP\ 0401 (2004) 003.




\bibitem{BMR5}
N.~Bartolo, S.~Matarrese and A.~Riotto, 
Phys.\ Rev.\ Lett.\  {\bf 93} (2004) 231301.

\bibitem{BMR6} N.~Bartolo, S.~Matarrese and A.~Riotto, astro-ph/0506410.

\bibitem{lythetal}  G.~I.~Rigopoulos, E.~P.~S.~Shellard and B.~W.~van Tent,
  astro-ph/0410486;
D.~H.~Lyth and Y.~Rodriguez,
  %``Non-gaussianity from the second-order cosmological perturbation,''
  Phys.\ Rev.\ D {\bf 71} (2005) 123508; {\it idem} astro-ph/0504045.


\bibitem{ks}
E.~Komatsu and D.~N.~Spergel,
%``Acoustic Signatures In The Primary Microwave Background Bispectrum,''
Phys.\ Rev.\ D 63 (2001) 063002.

\bibitem{k}
E.~Komatsu, et al., Astrophys.\ J.\ Suppl.\  148 (2003) 119.

\bibitem{sg} D.~N.~Spergel and D.~M.~Goldberg,
  %``Microwave background bispectrum. 1. Basic formalism,''
  Phys.\ Rev.\ D {\bf 59} (1999) 103001.

\bibitem{mg} A.~Gangui and J.~Martin,
  %``Best unbiased estimators for the three point correlators of the cosmic
  %microwave background radiation,''
  Phys.\ Rev.\ D {\bf 62}, 103004 (2000).


\bibitem{knox} L.~Knox,
  %``Determination of inflationary observables by cosmic microwave background
  %anisotropy experiments,''
  Phys.\ Rev.\ D {\bf 52}, 4307 (1995).


%\bibitem{curv1} D.~H.~Lyth, C.~Ungarelli and D.~Wands,
%  Phys. \ Rev. \ D {\bf 67} 023503 (2003)

%\bibitem{curv2} D.~H.~Lyth and D.~Wands,
%  Phys. \ Lett. B 524 (2002) 5

\bibitem{flatsky} F.~K.~Hansen, et al., in preparation 

\bibitem{Salopek} D.~S.~Salopek and J.~R.~Bond, Phys.\ Rev. \ D {\bf 42} 3936 (1990);
  {\em ibid.} {\bf 43} 1005 (1991)

\bibitem{Gangui} A.~Gangui, F.~Lucchin, S.~Matarrese and S.~Mollerach, Astrophys.\ J. 
  {\bf 430} 447 (1994)

\bibitem{Carroll} T.~Pyne and S.M.~Carroll, Phys.\ Rev. \ D {\bf 53} 2920 (1996) 

\bibitem{ksw} E.~Komatsu, D.N.~Spergel and B.D.~Wandelt,
Astrophys.~J., in press (astro-ph/0305189)

\bibitem{paolo} P.~Creminelli, A.~Nicolis, L.~Senatore, M.~Tegmark and
M.~Zaldarriaga, astro-ph/0509029

%%%%%%%%%%%%%%%%%%%%%%%%%%%%%%%%%%%%%%
%%%%%%%%%%%%%%%%%%%%%%%%%%%%%%%%%%%%%%
\end{thebibliography}
\end{document}